\documentclass[aps,twocolumn]{revtex4}
\usepackage{dcolumn}
\usepackage{graphicx}
\usepackage{amsmath}
\usepackage{amsfonts}
\usepackage{amssymb}
\usepackage{psfrag}
\usepackage{wrapfig}
\usepackage{subfigure}
\usepackage{makeidx}
\usepackage{bm}
\usepackage{epsf}
\usepackage{multirow}
\usepackage[colorlinks,urlcolor=blue,citecolor=blue]{hyperref}

\newcommand{\tabincell}[2]{\begin{tabular}{@{}#1@{}}#2\end{tabular}}

\begin{document}
\title{Classification of Dark Solitons via Topological Vector Potentials}
\author{L.-C. Zhao$^{1,2}$}
\author{Y.-H. Qin$^{1}$}
\author{J. Liu$^{3,4}$ }
\email{jliu@gscaep.ac.cn}
\address{$^{1}$School of Physics, Northwest University, Xi'an 710127, China}
\address{$^{2}$Shaanxi Key Laboratory for Theoretical Physics Frontiers, Xi'an 710127, China}
\address{$^{3}$Graduate School, China Academy of Engineering Physics, Beijing 100193, China}
\address{$^{4}$CAPT, HEDPS, and IFSA Collaborative Innovation Center of the Ministry of Education, Peking University, Beijing 100871, China}
 \begin{abstract}
Dark soliton is one of most interesting  nonlinear excitations in  physical systems, manifesting a spatially localized density ``dip" on  a uniform background accompanied  with a phase jump  across the dip.
However, the  topological properties of the dark solitons are far from fully understood.
Our investigation for the first time  uncover a vector potential underlying the nonlinear excitation
whose line integral gives the striking phase jump. More importantly, we find that the vector potential field has a topological configuration in analogous to  the Wess-Zumino term in a Lagrangian representation. It can induce  some  point-like magnetic  fields  scattered  periodically on a complex plane, each of them has a quantized magnetic  flux of elementary $\pi$.
We then calculate the Euler characteristic of the topological manifold of the  vector potential field  and
classify all known dark solitions according to the index.
\end{abstract}
\date{\today}

\maketitle

\emph{Introduction}---Dark soliton is one of most commonly nonlinear excitation emerged in both quantum and classical systems, including  optics \cite{Kivshar0,nop}, ultracold Bose-Einstein condensates \cite{Kevrekidis,BEC}, polariton fluid \cite{polaron1,polaron2,polaron3}, water waves \cite{Water} and the plasmas \cite{PM1,PM2}.
The generation of dark solitons are controllable and manipulatable in some situations \cite{Wu}, allowing for many important applications \cite{Anglin}, such as  optical communication, observation of negative mass effect \cite{PNAS},  atomic matter-wave interferometers \cite{matterinter},  quantum switches and splitters \cite{BEC}, etc..

Dark soliton serving as an exact solution for  nonlinear differential equations  has many implications in physics.
If we stand on a moving soliton to investigate its behavior, the dark soliton (or anti-dark soliton)
 will represent a kind of
transmission waves that can pass through a nonlinear potential well (or barrier for anti-dark soliton) without any reflection.
More interestingly, in contrast to its bright counterpart, there is a phase jump (or shift) during the process
 depending on the soliton's velocity. In the limit of zero velocity, i.e., for a stationary dark soliton, the phase jump usually takes $\pi$ value in many cases \cite{Kivshar0}.  It is thus reckoned  that the $\pi $ phase jump is  the signal of topological excitation and the stationary dark soltion is in several respects the one-dimensional counterpart to vortices \cite{Ku,Boris}. However, recent studies indicates that the phase jump can be greater than $\pi$ for a saturable nonlinear media \cite{Satuds,Satuds2},
for an anti-dark soliton it becomes zero and tend to $-\pi/2$ when the soliton velocity approaches sound speed \cite{CQds}.
Moreover, the phase jump for  the dark soliton in a derivative nonlinear system is found  to be $\pi/2$ that is independent of the soliton's velocity \cite{Chen}. Thus, the topological properties underlying the striking phase jumps  are far from fully understood.

On the other aspect, topology has emerged in real space
to manifest some important physical effects such as Aharonov-Bohm effect \cite{AB} or to demonstrate some  topological structures of  vortex \cite{Serani,Ku,Boris,Pismen}, skyrmions \cite{Donati,Savage},  and knots \cite{Knot,Hall}.
It can also emerge in parameter space such as Berry phase theory or momentum space such as topological energy band theory,
to reveal bizarre virtual  particles and characterize new forms of matter including topological insulators \cite{Hasan}, Weyl fermion semimetal \cite{Xu}, and even to promote the  quantum computing \cite{NA1,Rmp}.
In this paper, we investigate the topological properties of dark solitons in a complex space to  address the striking phase jumps.
We obtain an area theorem that can associate the phase jumps with an area on a plane of the amplitude vs local phase of a soliton solution. With exploiting  analytic extension of complex function, we uncover a topological vector potential
in analogous to  the Wess-Zumino term \cite{WZ}, whose line integral gives the striking phase jump.
The vector potential corresponds to some point-like magnetic  fields with  magnetic  flux of elementary $\pi$.
Our  result indicates  that, even though the dark soliton moving in real axis can not see any magnetic fields,  the  point-like magnetic  fields on the complex plane can actually affect the dark soliton's motion with assigning a phase variation.
We then  have made a topological classification for all known dark solitions according to their Euler characteristic of the  vector potential field.

\emph{Area theorem and topological vector potential}---A dark soliton is a spatially localized density ``dip" on top of a finite uniform background, accompanied with a phase jump through the dip \cite{Kivshar0,nop,Kevrekidis,BEC}. A finite phase step emerges at soliton center when the dark soliton is stationary. For a moving dark soliton,  there is a continuous phase shift across the soliton.  The solution for a dark soliton  can be depicted explicitly by a complex function $\psi(x,v,t)$,  where $x$ is the spatial coordinate, $v$ is the soliton's moving velocity, and $t$ is the evolution time. If we choose the soliton's  center as the reference to investigate the dark soliton's evolution, the  soliton  will be always stationary but its background admits a uniform density flow with a velocity $v$. In this frame, the dark soliton can be expressed as an eigenstate solution of form
$\bar\psi(x,v)$, with the boundary conditions:  $\lim\limits_{x \to -\infty} \bar\psi(x,v)=\sqrt{I} e^{i vx}$, and $\lim\limits_{x \to +\infty} \bar\psi(x,v)=\sqrt{I} e^{i vx+ i \Delta \phi_{ds}}$, where $I$ is the background density, $\Delta \phi_{ds}$ is the phase jump  across the dark soliton.
The local  phase of the  wave function $\bar\psi(x,v)$ can be written as
$\phi (x)=\phi_{ds}(x)+ vx$, where $v x$ term is from the extended  background flow and  the $\phi_{ds}(x)$ denotes the phase of the localized dark soliton solution. This analysis implies that a plane wave can propagate  from $-\infty$ to $+\infty $ without any reflection through  an effective quantum well induced by a dark soliton, the total phase jump can be expressed as
$\Delta \phi_{ds} = \int_{-\infty}^{+\infty} \frac{d\phi_{ds}(x)}{dx}dx.$

\begin{figure}[t]
\begin{center}
\includegraphics[height=40mm,width=85mm]{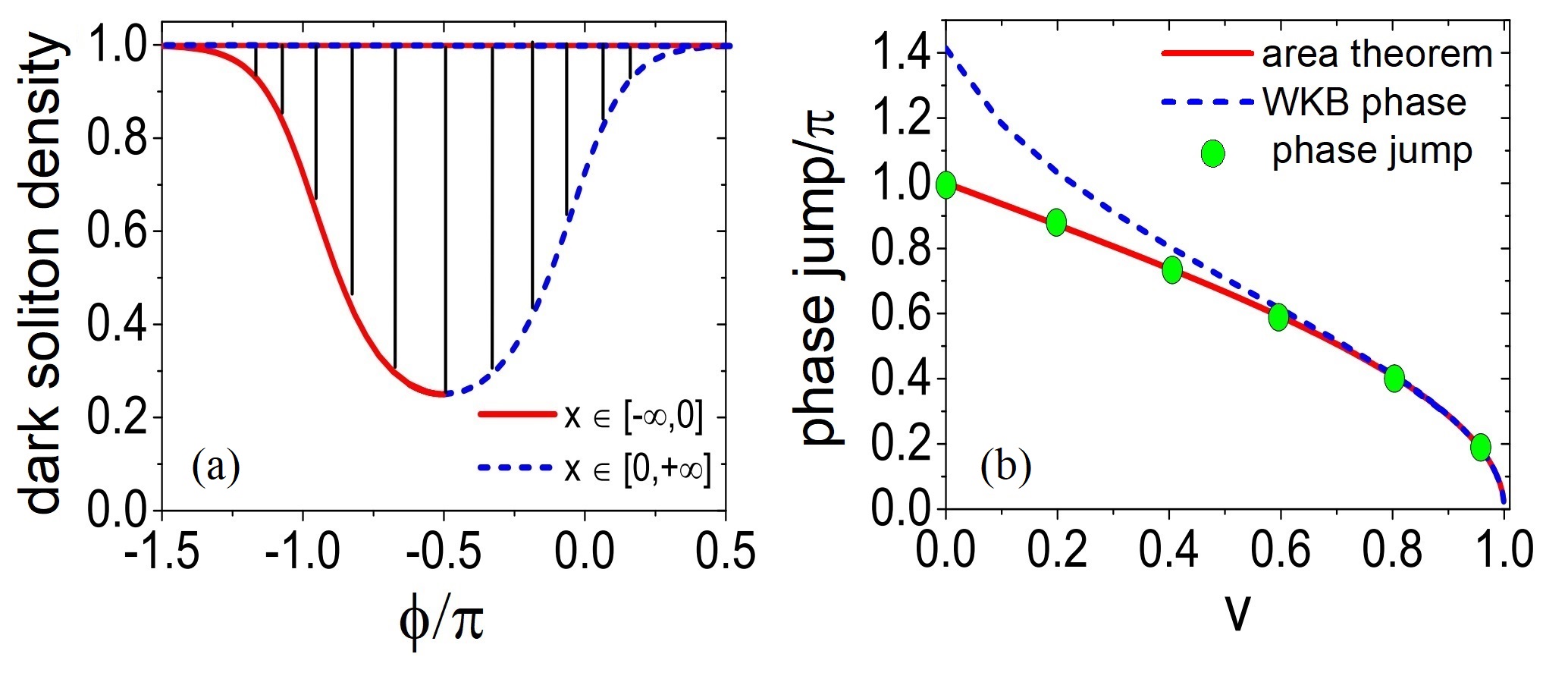}
\end{center}
\caption{(coloronline)(a) The area in the density-phase plane for a dark soliton solution. (b) The phase jump vs moving velocity for a scalar dark soliton. The red solid line is given by the area theorem,   blue dashed line is  WKB approximation,  and the green dots are  exact dark soliton solution. See text for details.  }\label{Fig1}
\end{figure}

The stationary wavefunction  $\bar\psi(x,v)$ can be viewed as an one-dimensional steady flow, satisfying the flow conservation of
 $|\bar\psi(x,v)|^2  \frac{d\phi(x)}{dx}=I v$. We  calculate
 $\int_{-\infty}^{+\infty} \bar\psi^*(-i\partial_x)\bar\psi dx =\int_{-\infty}^{+\infty} |\bar\psi(x,v)|^2  \frac{d\phi(x)}{dx} d x=\int_{-\infty}^{+\infty}Iv dx$, and then we have
 $\Delta \phi_{ds}
 =\int_{-\infty}^{+\infty} (\frac{d\phi(x)}{dx}-v ) dx
 =\int_{-\infty}^{+\infty} (1-|\bar\psi(x,v)|^2/{I}) d\phi$.
 The above formula implied that the the total phase shift of the dark soliton exactly corresponds to an area on the
 amplitude-phase plane. We term it as area theorem and sketch it in Fig.~\ref{Fig1} (a). Its validity has been verified by our numerical simulations as shown in Fig.~\ref{Fig1} (b). The area theorem is rather general,  not only  applicable to the simple scalar dark soltion (or anti-dark soliton) as discussed, but also  other  complicated vector solitons, such as dark-bright soliton \cite{DBS1,DBS2,DBS3}, spin soliton \cite{zhaoliu}, magnetic soliton \cite{Qu} and dark-bright-bright soliton \cite{DBB1}.  The area on the plane of amplitude vs. phase  in fact corresponds to  a classical canonical
action, which has a close relation to the Aharonov-Anandan phases of nonadiabatic evolutions \cite{Liu}.

 The area theorem can  help us to  uncover the topological properties of the dark solitons.
 We introduce a function $F[z]$, which is the analytic extension of  $F[x]=(1-|\bar\psi(x,v)|^2/I) {d\phi(x)}/dx$,
 with replacing $x$ by $z=x+i y$.  The phase jump of dark soliton can be  described by an integral of the vector potential $\textbf{A}$ along real $x$ coordinate,
 which  is introduced by considering a circle integral in the complex plane, i.e.,
$\oint_{C}  F[z] dz= \oint_{C} (u[x,y] + i w[x,y])  (dx +i dy)=\oint (u[x,y],-w[x,y])  \cdot d \textbf{r}+ i \oint (w[x,y],u[x,y])  \cdot  d \textbf{r}=\oint \textbf{A}  \cdot d \textbf{r}$,
where $\textbf{A}=u[x,y]\textbf{e}_x-w[x,y] \textbf{e}_y$, and $d\textbf{r}=dx\textbf{e}_x + dy\textbf{e}_y  $.
From the area theorem, we see that $F[z]$ might have  some  singularities of $z_N=x_N+i y_N$ ($N$ is an integer) on  complex plane corresponding to the divergence of flow velocity or the zero point of the density amplitude.
According to the Cauchy integral formula, a meromorphic function  can be expressed in terms of these singularities, that is,
 $F[z]= F(\infty)-\frac{1}{2\pi i}\int_\Gamma \frac{f(w)}{w-z}dw=\sum\limits_N \frac{\textbf{Res}[F[z_{N}]]}{z-z_{N}}$, where
 $\Gamma$ denotes the closed curves encircling the singularities separately and $ \textbf{Res}[F[z_{N}]]$ is the residue  \cite{LT}.
Our physical observation indicates that these singularities emerge periodically and in pairs on the complex plane,
which means that $\textbf{Res}[F[z_{N}]]$ can be expressed as $\pm \Omega/2\pi i $. Accordingly, the vector potential $\textbf{A}$ is derived as
\begin{eqnarray}
\textbf{A}&=& \sum_N \frac{\pm \Omega [(x-x_N) \textbf{e}_y-(y-y_N) \textbf{e}_x]} {2\pi [(x-x_N)^2+(y-y_N)^2]}.
\end{eqnarray}
 Here,  $\textbf{A}$ term takes the form of  Wess-Zumino topological term, i.e., a closed differential 1-form and  its differential 2-form is zero in the Lagrangian representation \cite{WZ}.
 This implies that
 corresponding magnetic filed will be  zero everywhere in whole complex  plane  except for those singular points.
  That is,
\begin{eqnarray}
\textbf{B}&=& \textbf{e}_z \sum_N \pm  \Omega \delta[\textbf{r}-\textbf{r}_N].
\end{eqnarray}
The $\pm$ represent the  magnetic flux direction.

It is interesting to compare with the Aharonov-Bohm effect \cite{AB} that predicts a topological phase when an electron moving on a close path around a solenoid.
A dark soliton solution moving on real axis can not see those magnetic fields scattered  on the complex plane, however it will acquire a phase jump due to the presence of  the  vector potential.
In  this sense, the phase jump for  the dark soliton can be viewed as 1D counterpart to the
famous Aharonov-Bohm phase.
However, the  phase shift  usually does not simply equal to the magnetic flux, because
the integral path for dark soliton is not a closed path around these singularities.
Moreover, we find that the flux has a quantized magnetic  flux of elementary $\Omega=\pi$,  corresponding to a monopole with a charge $1/2$ \cite{Liu3}.
In contrast to 2D topological excitation of vortex that admits a zero density core and a topological singularity of velocity field \cite{Serani,Ku,Boris,Pismen}, a moving dark soliton does not have a zero density core  and its fluid
field is continuous.

\begin{figure}[t]
\begin{center}
\includegraphics[height=120mm,width=85mm]{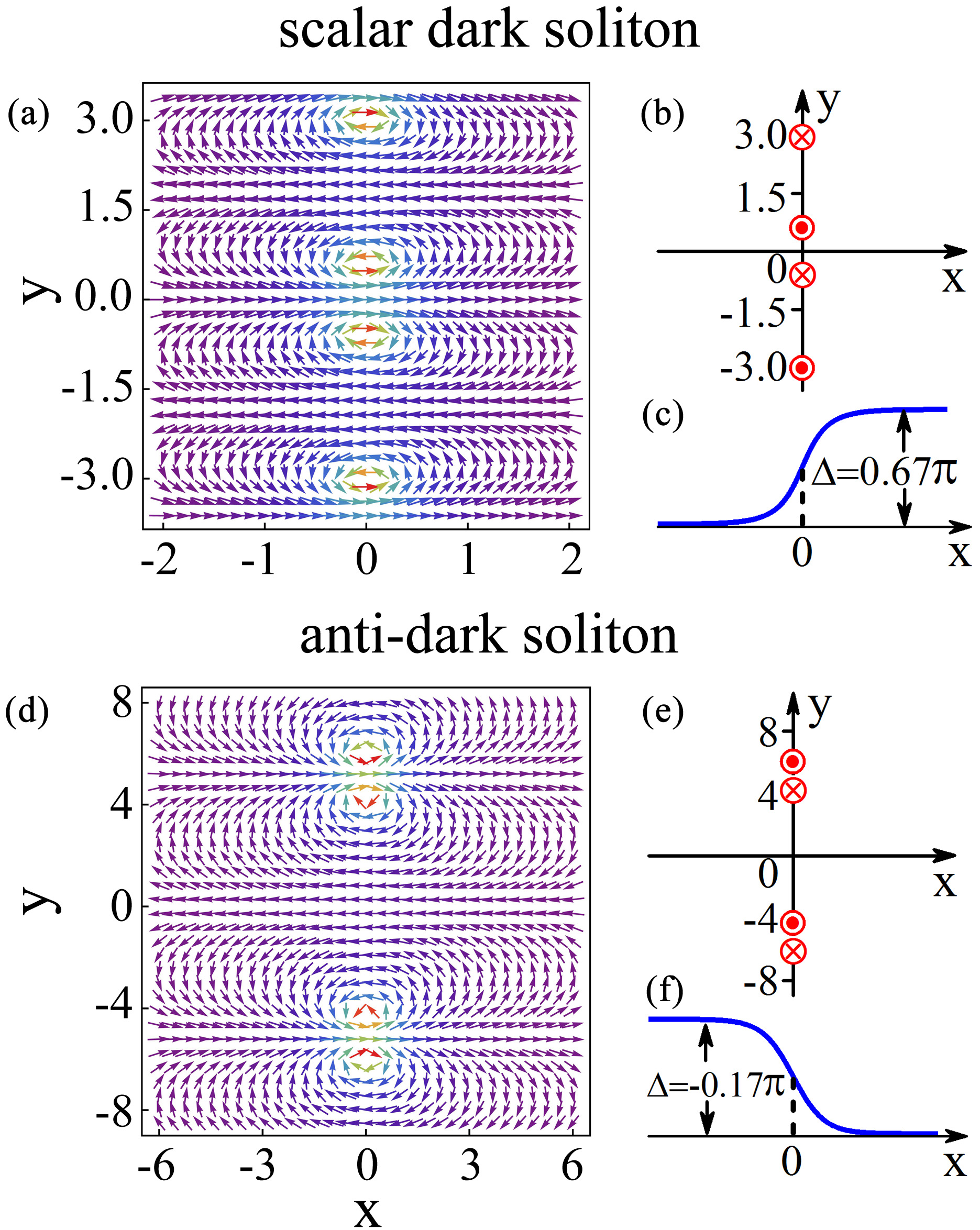}
\end{center}
\caption{ (a) The  topological vector potential $\textbf{A}$  for a scalar dark soliton.
(b) The corresponding magnetic field $\textbf{B}$ and (c) the phase jump induced by the magnetic field. The  soliton speed is set to be $v=0.5 c_s$.
(d) The vector potential $\textbf{A}$ for an anti-dark soliton \cite{CQds}.
 (e) The corresponding magnetic filed $\textbf{B}$ and (f) the phase jump induced by the magnetic field.  The soliton speed is set to be $v=0.5 c_s$. }\label{Fig2}
\end{figure}

\begin{figure}[t]
\begin{center}
\includegraphics[height=110mm,width=85mm]{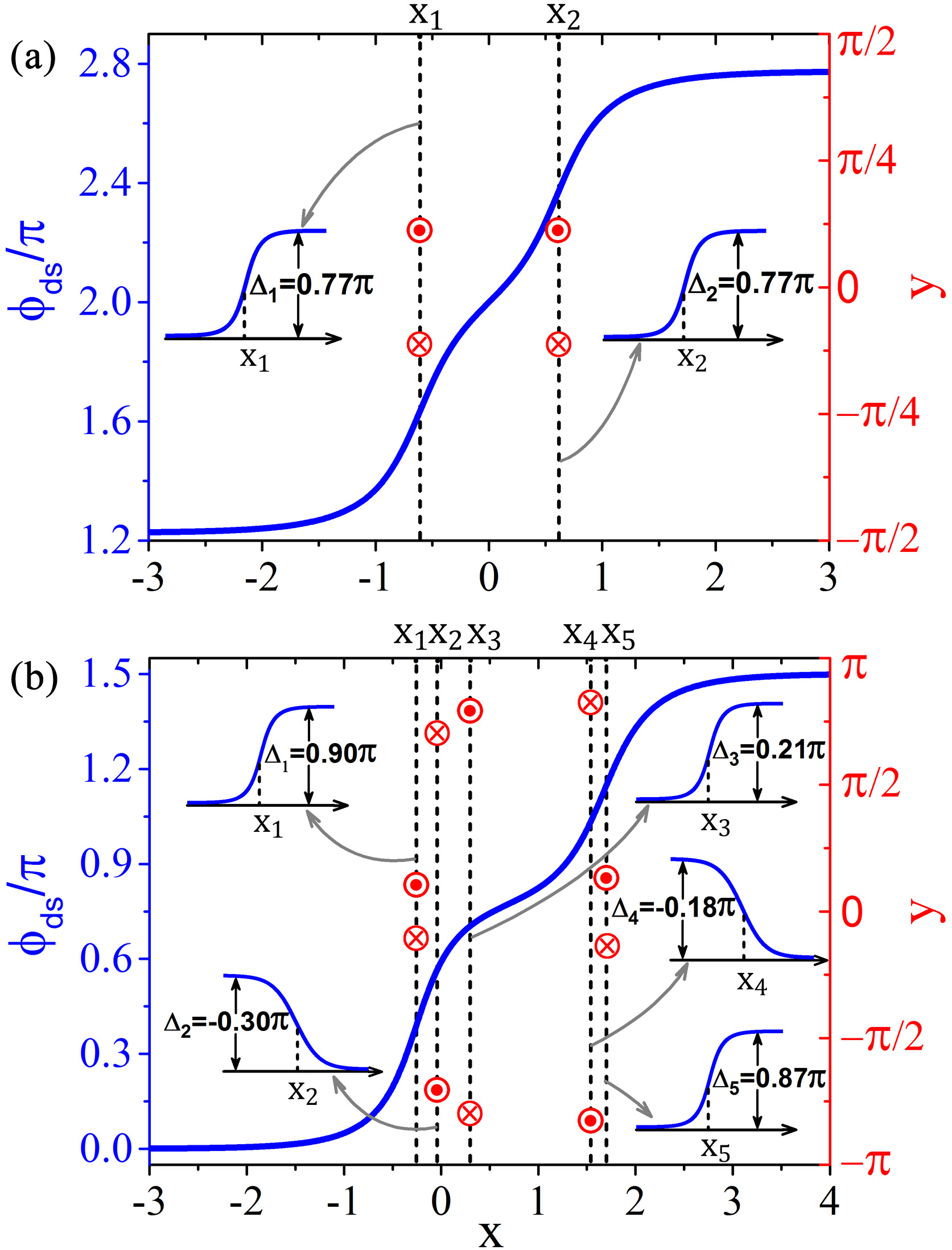}
\end{center}
\caption{ (a)  The phase distribution of the dark soliton phase and its corresponding magnetic filed $\textbf{B}$ on the complex plane for a dark-bright-bright soliton for which dark soliton admits symmetric double-valley. The soliton speed is $v=0.22 c_s$. The singularity locations are $(\pm 0.61, \pm 0.35)$  with a period of $\pi$ on the y-axis.
The  subgraphs show the phase variations induced by the singular magnetic fields on the two separate lines, respectively.  (b) The phase distribution of dark soliton and its corresponding magnetic filed $\textbf{B}$ for a dark-bright-bright soliton for which dark soliton admits asymmetric double-valley. The  singularity locations are $(-0.26, \pm  0.33) $, $ (- 0.04, \pm 2.21 )$, $(0.30, \pm 2.49 )$, $(1.54, \pm 2.59)$, and $(1.67, \pm 0.41)$  with a period of $2 \pi$ on the y-axis. The soliton speed is $v=0.19 c_s$. The  subgraphs show the phase variations induced by the singular magnetic fields on the five  separate lines, respectively. }\label{Fig3}
\end{figure}

\emph{Topological properties  of scalar dark solitons}---We first demonstrate our theory with a simple scalar soliton.
The scalar nonlinear  Schr\"odinger equation (NLS) of form
$ i \frac{\partial \psi}{\partial t}= -\frac{1}{2} \frac{\partial^2 \psi}{\partial x^2} + |\psi|^2 \psi$,  has wide applications in nonlinear fiber \cite{Kivshar0}, water wave \cite{Water}, plasma \cite{PM1}, and Bose-Einstein condensate \cite{BEC}.
It has a dark soliton solution with a uniform flowing background,
$\bar\psi(x,v)= (-i v+ \sqrt{1-v^2} \tanh [\sqrt{1-v^2} x] )  e^{i v x},$
with an eigenvalue of $\mu =1+v^2/2$. According to area theorem,
we can calculate the phase jump of the dark soliton from following integral expression,
$\Delta \phi_{ds}=\int_{-\infty}^{+\infty} (1-|\bar\psi|^2)  \frac{d \phi} {d x} dx
=\int_{-\infty}^{+\infty} \frac{ v \left(1-v^2\right)}{\cosh^2 \left(\sqrt{1-v^2} x\right)+ v^2-1} dx$.
The results are shown in Fig.~\ref{Fig1} (b). It indicates that the area theorem can precisely predicts the phase jump .

The phase variation of dark soliton can be also understood from the WKB approximation.
The amplitude of  dark soliton serves as  an  effective quantum well potential.
The classical  action is  $\int_{-\infty}^{+\infty} (\sqrt{2 k(x)}-v ) dx$, where $k(x)=\mu-|\bar\psi|^2$.
In WKB approximation, the  action should correspond to a quantum phase.   The results are   shown in Fig.~\ref{Fig1} (b). Interestingly,  the WKB phase agrees well with the area theorem in the limit when the soliton's velocity tends to sound speed,
while  they deviate dramatically in the low velocity  limit. This  can be understood from  the quantum-classical
correspondence \cite{Liu2,Liu3}.

We then can calculate the vector potential by determining  the singularity locations in the complex plane, i.e.,
\begin{eqnarray}
Z_N: x_N=0, y_N=\pm y_0 + N T , \,\,\,\,\,\,(N=0,\pm 1,...)
\end{eqnarray}
with $y_0=\frac{\arccos\left(\sqrt{1-v^2}\right)}{\sqrt{1-v^2}}$ and  $T=\frac{\pi}{\sqrt{1-v^2}}$ is the period.
The vector potential  is shown in Fig.~\ref{Fig2} (a). The corresponding magnetic filed $\textbf{B}=\nabla \times \textbf{A}$ can be obtained  accordingly,  which is shown in Fig.~\ref{Fig2} (b). The corresponding phase jump for dark soliton is shown in Fig.~\ref{Fig2} (c).
We see that the positive and negative magnetic flux emerge in pairs  and locate periodically  on the imaginary axis. Both the distance between the singularities and the period depends explicitly on the velocity of soliton. For instance,
in the limit of $v \to 0$, the magnetic fields at $\pm y_0$ tends to merge each other at origin.
With increasing the moving velocity to sound speed (here $c_s =1$), they tend to merge with other two magnetic flux
 at $y=\mp y_0\pm T$, respectively. In this limit, the dark soliton degenerates into a plane wave.

We now calculate the merge process when $v \to 0$ to understand $\pi$ jump of  the soliton  from   the vector potential perspective.
 When $v \rightarrow0$, $|y_0|\rightarrow 0$. In this limit,
\begin{eqnarray}
\lim_{v\rightarrow 0}\textbf{A}&=& \lim_{|y_0|\rightarrow 0} [\frac{\pi (x \textbf{e}_y+ |y_0| \textbf{e}_x)} {2\pi (x^2+ |y_0|^2)} + \frac{-\pi (x \textbf{e}_y- |y_0| \textbf{e}_x)} {2\pi (x^2+|y_0|^2)}] \nonumber\\
&=&\pi \delta[x] \textbf{e}_x .
\end{eqnarray}
Note that other singularities will merge each other in the process, and then  $\Delta \phi_{ds}= \int_{-\infty}^{+\infty} \textbf{A}_x dx= \pi$, indicating that the $\pi$ jump arises from the  quantized magnetic  flux  corresponding to a monopole.

The phase jump for the anti-dark soliton \cite{CQds} is also of interest. It is zero for the static solution  and  tend to be $-\pi/2$ when the moving speed approaching sound speed.
This behavior can be explained by underlying topological potential. The vector potentials and magnetic fields are shown Fig.~\ref{Fig2} (d) and (e).
In the limit of $v \to 0$, the paired  magnetic fields tends to merge each other and gives zero phase jump. With increasing the soliton velocity to sound speed, they tend to locate at imaginary axis separately.
Comparing Fig.~\ref{Fig2} (e) with the  Fig.~\ref{Fig2} (b), we find  that the directions of the corresponding flux
 are reversed that  leads to a negative phase jump (see Fig.~\ref{Fig2} (f)).

\begin{table}[t]
\small
\centering
\begin{tabular}{|c|c|c|c|c|c|}
\hline
Types  &   $\Delta \phi_{ds}$  & T & g & M   &$\chi$ \\
\hline
 DS \cite{Kivshar0}  & $[0,\pi]$ & \multirow{4}*{$\frac{\pi}{\sqrt{c_s^2-v^2}}$ } & \multirow{8}*{ 2} & \multirow{8}*{$S^1\times R^1/Z_0Z_1$} & \multirow{8}*{-4} \\
 \cline{1-2}
 DS \cite{CQds} &  $[\frac{\pi}{2},\pi]$ &  &   &      & \\
\cline{1-2}
  A-DS \cite{CQds} &  $[-\frac{\pi}{2},0]$ & &   &     & \\
  \cline{1-2}
  SS \cite{zhaoliu} &  $[0,\pi]$ &  &   &      & \\
\cline{1-3}
 DS \cite{Qds} &   \tabincell{c}{$[0,\pi]$} & \tabincell{c}{$\frac{2\pi}{\sqrt{c_s^2-v^2}}$} &  &      & \\
 \cline{1-3}
 DS \cite{Chen} &   $\frac{\pi}{2}$ & $\frac{4\pi}{v}$ &  &      & \\
\cline{1-3}
 DBS \cite{DBS3}  &   $[0,\pi]$ & $\pi$ &   &   &  \\
 \cline{1-3}
 DBBS \cite{DBB1} &   $[0,\pi]$ & $\pi$ &  &      & \\

 \hline
  DS \cite{Satuds,Satuds2} &   $[0,2\pi]$  & -- & -- & --   &-- \\
   \hline
  MS \cite{Qu} &  \tabincell{c}{$[\frac{\pi}{2},\pi]$ \\ $[-\frac{\pi}{2},0]$} & --  & --  &--  &--\\
 \hline
DBBS1  &   $[0,2\pi]$ & $\pi$ &4 &$S^1\times R^1/Z_0...Z_3$  &-8\\
\hline
DBBS2  &  $[0,2\pi]$ & $2\pi$ & 10 &$S^1\times R^1/Z_0...Z_{9}$   &-20\\
 \hline
DBBBS  &   $[0,3\pi]$ & $\pi$ & 6 & $S^1\times R^1/Z_0...Z_5$ &-12\\
\hline
\end{tabular}
\caption{ $\Delta \phi_{ds}$ the interval of phase jump;
$T$ is the period in $y$ axis where
$c_s$ is the sound speed;
$g$ refers to the number of the singularities in one period;
$M$ is  topological manifold space;
$Z_N$ represents singular points;
$\chi=0-2g$ is the Euler characteristic number.
The solutions of  DBBS1, DBBS2 and  DBBBS  are obtained from present work\cite{supplm}.
 For MS \cite{Qu}, the phase jump regimes $[\frac{\pi}{2},\pi]$ and $[-\frac{\pi}{2},0]$ correspond to dark and anti-dark soliton components, respectively. `--' means corresponding quantities can not be calculated due to the absence of exact explicit expressions.  }
\end{table}

\emph{Topological properties  of vector dark solitons}---
We now extend our discussions to a more complicated three-component coupled NLS system. With using the developed Darbox transformation \cite{Ling,Qin} and after lengthy deductions, we obtain a kind of dark-bright-bright soliton (DBBS)\cite{supplm}, for which there is a dark soliton with a double-valley density profile in one component, and  bright soliton density profiles in the other two components. Interestingly, the phase jump for this dark soliton component is in the regime $[0,2\pi]$.
 The double-valley structure might be  symmetric or asymmetric.
Corresponding magnetic fields and the dark soliton phase $\phi_{ds}$ are calculated and plotted  in Fig.~\ref{Fig3}, respectively.
 For the symmetric one, the singular points scattered on  two separate lines with $x=x_1$ and $x=x_2$
 with a period of $\pi$ along $y$ axis. The phase variation induced by the point-like magnetic fields periodically scattered
 on two separate lines is found to be $\Delta_1=\Delta_2= 0.77\pi$, sum of them gives total phase jump of the dark soliton.   While for the asymmetric one, the singularities locate on  five separate lines at $x_j$ ($j=1,2,3,4,5$),
 and the period along $y$ axis turns to be  $2 \pi$.
 We also calculate the  phase variations $\Delta_j$ ($j=1,2,3,4,5$) corresponding to the magnetic fields on the  five separate  lines and show the results in the  subgraphs of  Fig.~\ref{Fig3} (b).
 The phase profiles in Fig.~\ref{Fig3}  demonstrate multi-steps structures due to the complicated  topological potentials.

\emph{Topological classification of the dark solitons}---
We collect all known dark soliton solutions and
calculate  their corresponding vector potential fields and the Euler characteristic  of the topological manifolds. The results are presented in Table I.  We find that that all previous dark soliton solutions correspond to Euler index of $-4$, while the complicated vector solitons obtained  in the present work have the higher topological number up to $-20$.

From Table I, we see that the dark soliton in a derivative nonlinear system is very special, in which the phase jump is $\frac{\pi}{2}$  independent of the soliton velocity \cite{Chen}.  However, the singularities of vector potential locate at $\pm i \frac{\pi}{v} + i N T$ with $T=4\pi/v$, whose distribution obviously  depends  on the velocity.
Nevertheless,  after a simple scaling transformation,  an equivalent topological vector potential can be derived in a  velocity independent form, i.e.,  $ \bar{\textbf{A}}=\sum\limits_N \frac{\pm  [ x \textbf{e}_y-(y-\bar y_N) \textbf{e}_x]} {2 [ x^2+(y-\bar y_N)^2]} $ with the singularities of $\bar y_N=\pm \frac{\pi}{2} +2\pi N  $, whose line integral will gives a definite  $\pi/2$ phase jump.

\emph{Conclusion}---
We show that the  dark solitons serving as a kind of simple nonlinear excitations can demonstrate very interesting topological properties. The  phase jump can be viewed as the 1D counterpart to the famous  Aharonov-Bohm phase,
where the  solenoid that might carry an arbitrary flux is replaced by a monopole with  a quantized magnetic  flux of elementary $\pi$.
Underlying vector fields demonstrate the topology of Wess-Zumino term. Our investigations have resolved a long-standing puzzle on  the topological origin of dark solitons and  provides a possibility to investigate topological vector potential via the generation of dark solitons that are controllable in current BEC, optic and microcavity polariton condensates experiments.


\begin{thebibliography}{99}
\bibitem{Kivshar0} Y.S. Kivshar and B. Luther-Davies, \newblock Dark optical solitons: physics and applications,  \newblock
    \href{https://doi.org/10.1016/S0370-1573(97)00073-2}  {Phys. Rep. \textbf{298}, 81 (1998).}
\bibitem{nop} Y.S. Kivshar and G.P. Agrawal, \emph{Optical Solitons: From Fibers to Photonic Crystals} (Academic, NewYork, 2003).
\bibitem{Kevrekidis} P.G. Kevrekidis, D.J. Frantzeskakis, and R. Carretero-Gonz\'{a}lez, \emph{Emergent Nonlinear Phenomena in Bose-Einstein Condensates: Theory and Experiment} (Springer, Berlin Heidelberg, 2008).
\bibitem{BEC} D.J. Frantzeskakis, Dark solitons in atomic Bose-Einstein condensates: from theory to experiments,  \newblock
    \href{https://doi.org/10.1088/1751-8113/43/21/213001} { J. Phys. A: Math. Theor. \textbf{43}, 213001 (2010).}
\bibitem{polaron1} Y. Xue and M. Matuszewski, Creation and Abrupt Decay of a Quasistationary Dark Soliton in a Polariton Condensate, \newblock
    \href{https://doi.org/10.1103/PhysRevLett.112.216401}  {Phys. Rev. Lett. \textbf{112}, 216401 (2014).}
\bibitem{polaron2} V. Goblot, H.S. Nguyen, I. Carusotto, E. Galopin, A. Lema\^{\i}tre, I. Sagnes, A. Amo, and J. Bloch, Phase-Controlled Bistability of a Dark Soliton Train in a Polariton Fluid, \newblock
    \href{https://doi.org/10.1103/PhysRevLett.117.217401}  {Phys. Rev. Lett. \textbf{117}, 217401 (2016).}
\bibitem{polaron3} X. Ma, O.A. Egorov, and S. Schumacher, Creation and Manipulation of Stable Dark Solitons and Vortices in Microcavity Polariton Condensates, \newblock
    \href{https://doi.org/10.1103/PhysRevLett.118.157401}  {Phys. Rev. Lett. \textbf{118}, 157401 (2017).}

\bibitem{Water} A. Chabchoub, O. Kimmoun, H. Branger, N. Hoffmann, D. Proment, M. Onorato, and N. Akhmediev, Experimental Observation of Dark Solitons on the Surface of Water, \newblock
    \href{https://doi.org/10.1103/PhysRevLett.110.124101}  {Phys. Rev. Lett. \textbf{110}, 124101 (2013).}

\bibitem{PM1} P.K. Shukla and B. Eliasson, Formation and Dynamics of Dark Solitons and Vortices in Quantum Electron Plasmas,  \newblock
    \href{https://doi.org/10.1103/PhysRevLett.96.245001}  {Phys. Rev. Lett. \textbf{96}, 245001 (2006).}
\bibitem{PM2} R. Heidemann, S. Zhdanov, R. S\"{u}tterlin,  H.M. Thomas, and G.E. Morfill, Dissipative Dark Soliton in a Complex Plasma, \newblock
    \href{https://doi.org/10.1103/PhysRevLett.102.135002}  { Phys. Rev. Lett. \textbf{102}, 135002 (2009).}
\bibitem{Wu} B. Wu, J. Liu, and Q. Niu, Controlled Generation of Dark Solitons with Phase Imprinting,
\newblock \href{https://doi.org/10.1103/PhysRevLett.88.034101}  { Phys. Rev. Lett. \textbf{88}, 034101 (2002).}
\bibitem{Anglin} J. Anglin, Atomic dark solitons: Quantum canaries learn to fly,  \newblock
    \href{https://doi.org/10.1038/nphys980} {Nature Phys. \textbf{4}, 437 (2008).}

\bibitem{PNAS} L.M. Aycock, H.M. Hurst, D.K. Efimkin, D. Genkina, H.-I. Lu, V.M. Galitski,
and I.B. Spielman, \newblock Brownian motion of solitons in a Bose-Einstein condensate, \newblock \href{https://doi.org/10.1073/pnas.1615004114}
    {PNAS \textbf{114}, 2503 (2017).}
\bibitem{matterinter} C. Lee, E.A. Ostrovskaya and Y.S. Kivshar, Nonlinearity-assisted quantum tunnelling in a matter-wave interferometer, \newblock \href{https://doi.org/10.1088/0953-4075/40/21/010} {J. Phys. B: At. Mol. Opt. Phys. \textbf{40}, 4235 (2007).}
\bibitem{Ku} M.J.H. Ku, W. Ji, B. Mukherjee, E. Guardado-Sanchez, L.W. Cheuk, T. Yefsah, and M.W. Zwierlein, Motion of a Solitonic Vortex in the BEC-BCS Crossover,   \newblock
    \href{https://doi.org/10.1103/PhysRevLett.113.065301}  {Phys. Rev. Lett. \textbf{113}, 065301 (2014).}
\bibitem{Boris} B.A. Malomed, Vortex solitons: Old results and new perspectives,  \newblock
    \href{https://doi.org/10.1016/j.physd.2019.04.009}  {Physica D \textbf{399}, 108 (2019).}
\bibitem{Satuds} W. Krolikowski and B. Luther-Davies, Dark optical solitons in saturable nonlinear media, \newblock
    \href{https://doi.org/10.1364/OL.18.000188} {Opt. Lett. \textbf{18}, 188 (1993).}
\bibitem{Satuds2} W. Kr\'{o}likowski, N. Akhmediev, and B. Luther-Davies, Darker-than-black solitons: Dark solitons with total phase shift greater than $\pi$, \newblock
    \href{https://doi.org/10.1103/PhysRevE.48.3980} {Phys. Rev. E \textbf{48}, 3980 (1993).}

\bibitem{CQds} Y.S. Kivshar, V.V. Afansjev, and A.W. Snyder, Dark-like bright solitons, \newblock
    \href{https://doi.org/10.1016/0030-4018(96)00111-3}  {Opt. Commun. \textbf{126}, 348 (1996).} The related parameters are chosen as $\alpha=\frac{3}{8}$, $I_{\infty}=1$ for the analysis.

\bibitem{Chen} H. Triki, Y. Hamaizi, Q. Zhou, A. Biswas, M.Z. Ullah, S.P. Moshokoa, and M. Belic, Chirped dark and gray solitons for Chen-Lee-Liu equation in optical fibers and PCF, \newblock
    \href{https://doi.org/10.1016/j.ijleo.2017.11.038}  {Optik \textbf{155}, 329 (2018).} We choose the simple Case-I dark soliton solution with $a=b=1$ for analysis.

\bibitem{AB} Y. Aharonov and D. Bohm, Significance of Electromagnetic Potentials in the Quantum Theory, \newblock
    \href{https://doi.org/10.1103/PhysRev.115.485} {Phys. Rev. \textbf{115}, 485 (1959).}
\bibitem{Pismen}  L.M. Pismen, \emph{Vortices in Nonlinear fields} (Oxford, Clarendon, 1999).
\bibitem{Serani} S. Donadello, S. Serafini, M. Tylutki, L.P. Pitaevskii, F. Dalfovo, G. Lamporesi, and G. Ferrari, Observation of Solitonic Vortices in Bose-Einstein Condensates,  \newblock
    \href{https://doi.org/10.1103/PhysRevLett.113.065302}  {Phys. Rev. Lett. \textbf{113}, 065302 (2014).}

\bibitem{Donati} S. Donati, L. Dominici, G. Dagvadorj, D. Ballarini, M.D. Giorgi, A. Bramati, G. Gigli, Y.G. Rubo, M.H. Szyma\'{n}ska, and D. Sanvitto, Twist of generalized skyrmions and spin vortices in a polariton superfluid, \newblock
    \href{https://doi.org/10.1073/pnas.1610123114} {PNAS \textbf{113}, 14926 (2016).}
\bibitem{Savage} C.M. Savage and J. Ruostekoski, Energetically Stable Particlelike Skyrmions in a Trapped Bose-Einstein Condensate,  \newblock
    \href{https://doi.org/10.1103/PhysRevLett.91.010403} {Phys. Rev. Lett. \textbf{91}, 010403 (2003).}
\bibitem{Knot} M. Atiyah, \emph{The geometry and physics of knots} (Cambridge University Press, 1990).
\bibitem{Hall} D.S. Hall, M.W. Ray, K. Tiurev, E. Ruokokoski, A.H. Gheorghe, and M. M\"{o}tt\"{o}nen, Tying quantum knots, \newblock
    \href{https://doi.org/10.1038/nphys3624}  {Nature Phys. \textbf{12}, 478 (2016).}
\bibitem{Hasan} M.Z. Hasan and C.L. Kane, Colloquium: Topological insulators, \newblock
    \href{https://doi.org/10.1103/RevModPhys.82.3045}  {Rev. Mod. Phys. \textbf{82}, 3045 (2010).}

\bibitem{Xu} S.-Y. Xu, I. Belopolski, N. Alidoust, M. Neupane, et al., Discovery of a Weyl fermion semimetal and topological Fermi arcs,  \newblock
    \href{https://doi.org/10.1126/science.aaa9297}
 {Science \textbf{349}, 613 (2015).}
\bibitem{NA1} E. Gibney and D. Castelvecchi, Physics of 2D exotic matter wins Nobel,  \newblock
    \href{https://doi.org/10.1038/nature.2016.20722} {Nature \textbf{538}, 18 (2016).}
\bibitem{Rmp} D. Castelvecchi, The strange topology that is reshaping physics,  \newblock
    \href{https://doi.org/10.1038/547272a} {Nature \textbf{547}, 272 (2017).}
 \bibitem{WZ}  J. Wess and B. Zumino, Consequences of anomalous ward identities, \newblock
    \href{https://doi.org/10.1016/0370-2693(71)90582-X} {Phys. Lett. B \textbf{37}, 95 (1971).}

\bibitem{DBS1} S. Trillo, S. Wabnitz, E.M. Wright, and G.I. Stegeman, Optical solitary waves induced by cross-phase modulation,  \newblock
    \href{https://doi.org/10.1364/OL.13.000871} {Opt. Lett. \textbf{13}, 871 (1988).}
\bibitem{DBS2} D.N. Christodoulides, Black and white vector solitons in weakly birefringent optical fibers, \newblock
    \href{https://doi.org/10.1016/0375-9601(88)90511-7} {Phys. Lett. A \textbf{132}, 451 (1988).}
\bibitem{DBS3} Th. Busch and J.R. Anglin, Dark-Bright Solitons in Inhomogeneous Bose-Einstein Condensates, \newblock
    \href{https://doi.org/10.1103/PhysRevLett.87.010401} {Phys. Rev. Lett. \textbf{87}, 010401 (2001).} The parameter $\kappa=1$ is set  for our analysis.
\bibitem{zhaoliu}  L.-C. Zhao, W. Wang, Q. Tang, Z.-Y. Yang, W.-L. Yang, and J. Liu,
 \newblock Spin soliton with a negative-positive mass transition, \newblock
 \href{https://doi.org/10.1103/PhysRevA.101.043621}%
 {Phys. Rev. A \textbf{101}, 043621 (2020).}
\bibitem{Qu} C. Qu, L.P. Pitaevskii, and S. Stringari,
 \newblock Magnetic solitons in a Binary Bose-Einstein Condensate, \newblock
 \href{https://doi.org/10.1103/PhysRevLett.116.160402}%
 {Phys. Rev. Lett. \textbf{116}, 160402 (2016).}
 \bibitem{DBB1} R. Radhakrishnan, N. Manikandan, and K. Aravinthan, Energy-exchange collisions of dark-bright-bright vector solitons, \newblock
    \href{https://doi.org/10.1103/PhysRevE.92.062913} {Phys. Rev. E \textbf{92}, 062913 (2015).} The parameter $k_R^{(1)}=1$ is set for analysis.

\bibitem{Liu} J. Liu, B. Wu, and Q. Niu, Nonlinear Evolution of Quantum States in the Adiabatic Regime, \newblock
    \href{https://doi.org/10.1103/PhysRevLett.90.170404}  {Phys. Rev. Lett. \textbf{90}, 170404 (2003).}
\bibitem{LT} W.K. Hayman, \emph{Meromorphic Functions} (Oxford University Press, 1964).
\bibitem{Liu3} J. Liu, S.-C. Li, L.-B. Fu, and D.-F. Ye, \emph{Nonlinear Adiabatic Evolution of Quantum Systems } (Springer Nature Singapore Pte Ltd, 2018).
\bibitem{Liu2} B. Wu and J. Liu, Commutability between the Semiclassical and Adiabatic Limits, \newblock
    \href{https://doi.org/10.1103/PhysRevLett.96.020405} {Phys. Rev. Lett. \textbf{96}, 020405 (2006).}

\bibitem{Ling} L. Ling, L.-C. Zhao, and B. Guo, Darboux transformation and multi-dark
soliton for N-component nonlinear
Schr\"odinger  equations,  \newblock
    \href{https://doi.org/10.1088/0951-7715/28/9/3243}%
    {Nonlinearity \textbf{28}, 3243 (2015).}
\bibitem{Qin} Y.-H. Qin, L.-C. Zhao, and L. Ling, Nondegenerate bound-state solitons in multicomponent Bose-Einstein condensates, \newblock
    \href{https://doi.org/10.1103/PhysRevE.100.022212} {Phys. Rev. E \textbf{100}, 022212 (2019).}

\bibitem{supplm} The Supplemental material for this paper.

\bibitem{Qds} M.A. Alejo and  A.J. Corcho, Orbital stability of the black soliton for the quintic Gross-Pitaevskii equation, \newblock
    \href{https://arxiv.org/abs/2003.09994} {arXiv:2003.09994 (2020).}
 \end{thebibliography}
\end{document}